\documentclass{appolb}
\usepackage{epsfig}
\usepackage{axodraw}
\def\ra{\rightarrow}
\def\k{m_{eff}}
\def\e{\epsilon}
\def\p{\pi}
\def\th{\theta}

\def\bar{\overline}

\newcommand{\AmS}{{\protect\the\textfont2
  A\kern-.1667em\lower.5ex\hbox{M}\kern-.125emS}}

\def\beq{ \begin{equation}}
\def\eeq{\end{equation} }
\def\bea{\begin{eqnarray}}
\def\eea{\end{eqnarray}}

\if@twoside
\else
\fi
\if@twocolumn
  \twocolumn
\else
  \onecolumn
\fi

\def\runhead{}

\begin{document}

% \eqsec  % uncomment this line to get equations numbered by (sec.num)

\title{
\vspace*{-4 cm}
\begin{flushright}
hep-ph/0005093 \\
CERN-TH-2000-137 \\
\end{flushright}
\vspace*{2 cm}
{\bf RENORMALISATION EFFECTS OF NEUTRINO \\
MASSES AND INTERACTIONS}~\thanks{Invited talk at the
Cracow Epiphany Conference on Neutrinos in Physics and Astrophysics, 
January 2000}%
% you can use '\\' to break lines
}
\author{\bf Smaragda Lola
\address{CERN Theory Division, CH-1211, Geneva 23, Switzerland}
}
\maketitle
\begin{abstract}
{\bf \tt \bf Contents:}

\hspace*{0.5 cm}
{\bf \tt \bf 1.}
Data and implications.

\hspace*{0.5 cm}
{\bf \tt \bf 2.}
Neutrino threshold effects.

\hspace*{0.5 cm}
{\bf \tt \bf 3.}
Renormalisation of the neutrino mass operator
and stability properties of neutrino textures.

\hspace*{0.5 cm}
{\bf \tt \bf 4.}
Neutrino threshold effects  and Yukawa
unification.

\hspace*{0.5 cm}
{\bf \tt \bf 5.}
Renormalisation-induced lepton-flavour-violating processes
from non-zero neutrino masses.

\hspace*{0.5 cm}
{\bf \tt \bf 6.}
Summary.
\end{abstract}


\section{Data and implications}

\noindent
The Super-Kamiokande data \cite{SKam}
clearly indicate a $\nu _{\mu }/\nu _{e }$ ratio
in the atmosphere that is significantly 
smaller than the theoretical expectations.
The most natural way to explain this deviation is by
introducing $\nu _{\mu }$--$\nu _{\tau }$ oscillations, with 
$\delta m_{\nu _{\mu }\nu _{\tau }}^{2} \approx (10^{-2}\;{\rm to}%
\;10^{-3})\;{\rm eV^{2}}$ and 
$\sin^{2}2\theta _{\mu \tau } \geq 0.85$.
Alternative schemes with dominant
$\nu _{\mu }\rightarrow \nu _{e}$
oscillations are excluded by both 
the  Super-Kamiokande data on electron-like events~\cite{SKam}, 
and the Chooz reactor 
experiment~\cite{chooz}.
Finally, oscillations involving a sterile neutrino
are disfavoured (but not yet excluded) by 
the azimuthal-angle dependence of muon-like events~\cite{SKam}
and by measurements of $\pi^0$ production.

Once neutrino oscillations are introduced in order to
explain the atmospheric neutrino deficit, it is natural
to address similarly the solar neutrino puzzle.
The latter  can be resolved through
either
vacuum or matter-enhanced (MSW) oscillations. The first require a mass
splitting of the neutrinos
that are involved in the oscillations 
in the range
$\delta m_{\nu _{e}\nu _{\alpha }}^{2} 
\sim (0.5-1.1)\times 10^{-10}~{\rm eV}^2$,
where $\alpha $ is $\mu $ or $\tau$. 
MSW oscillations~\cite{MSW}, on the other hand, require
$ \delta m_{\nu _{e}\nu _{\alpha }}^{2} \sim (0.3-20) \times 10^{-5}
~{\rm eV}^2$ with either large $\sin^{2}2\theta _{e \alpha } \sim 1$
or small $\sin^{2}2\theta _{e \alpha} \sim 10^{-2}$.

The implications of these observations are very interesting, since
they  point towards a non-zero neutrino mass and lepton-number 
violation, that is
{\em the existence of physics  beyond
the standard model}. 
It turns out that both the solar and atmospheric neutrino
data can be accommodated in a natural way
in schemes with three light
neutrinos with at least one large mixing angle and hierarchical
masses, of the order of the required
mass differences:
$m_{3} \sim (10^{-1} \; {\rm to} \;
10^{-1.5})$ eV
and $m_{2} \sim (10^{-2} \; {\rm to} \;
10^{-3})$ eV $\gg m_3$.
On the other hand,
if neutrinos are also to provide a
significant hot dark matter component,
three almost-degenerate
neutrinos with  masses of $\approx 1$ eV
would be required.

Along these lines, a natural question that arises is why neutrino
masses are smaller that the rest of the
fermion masses in the theory. 
This can be explained by the see-saw mechanism \cite{seesaw}, which
involves Dirac neutrino masses $m_{\nu}^D$, 
of the same order as
the charged-lepton and quark masses, and heavy Majorana
masses $M_{\nu_R}$ for the right-handed
neutrinos, $\nu_R$,
in a way that light effective neutrino mass
matrices at a scale $M_N$, such that:
\begin{equation}
m_{eff}
=m^D_{\nu}\cdot (M_{\nu_R})^{-1}\cdot m^{D^{T}}_{\nu}
\label{eq:meff}
\end{equation}
For instance, for {$m_{\nu}^D \approx 200 ~ {\rm GeV}$}
and 
$M_N \approx {\cal O}({\rm 10^{13}} ~ {\rm GeV})$, 
{ $m_{eff} \approx$ 1~ eV}.
Then,  in complete analogy to the quark
currents, the leptonic mixing matrix is \cite{mns}
\bea
V_{MNS} =
V_{\ell}V_{\nu}^{\dagger}
\eea
where $V_{\ell}$ diagonalizes the charged-lepton mass matrix, while
$V_{\nu}$ diagonalizes the light neutrino mass matrix $m_{eff}$.

In the presence of neutrino masses,
the running of the various couplings
from the unification scale
down to low energies is modified.
From $M_{GUT}$ to $M_N$, one  must include
 radiative corrections  from the neutrino Yukawa couplings, 
while below $M_N$, the right-handed neutrinos decouple from the 
spectrum and an effective  see-saw mechanism is operative.
It actually turns out, as we are going to
discuss in subsequent sections, that the renormalisation 
effects give important information on
the structure of the neutrino textures, 
while
unification can also be affected by
neutrino thresholds.

Neutrino oscillations involve violations of the
individual lepton numbers $L_{e, \mu, \tau}$,
raising the prospect that there might also exist
observable processes that violate 
charged-lepton-number conservation \cite{neutrinoLFVns,
neutrinoLFVs}, such as 
$\mu \rightarrow e \gamma$, $\mu \rightarrow 3 e$,
$\tau \rightarrow \mu \gamma$,  and $\mu-e$
conversion on heavy nuclei~\cite{neutrinoLFVs,a,KO}.
In non-supersymmetric models with
massive neutrinos, the amplitudes for the
charged-lepton-flavour violation are 
proportional to inverse powers of
the right-handed neutrino mass scale 
$M_{\nu_R}$, and thus
the rates for rare decays are extremely
suppressed~\cite{neutrinoLFVns}.
On the other hand, in supersymmetric models these
processes are only suppressed
by inverse powers of the supersymmetry breaking scale, which is 
at most $1$~TeV \cite{neutrinoLFVs}.
The present experimental upper limits on the most
interesting of these decays are
\begin{eqnarray}
BR(\mu \ra e \gamma) \, < \, 1.2 \times 10^{-11} \, && \, \,
\cite{Brooks} \\
BR(\mu^+ \ra e^+ e^+ e^-) \, < \, 1.0 \times 10^{-12} \, && \, \,
\cite{Bellgardt} \\
R(\mu^- Ti \ra e^- Ti) \, < \, 6.1 \times 10^{-13} \, && \, \,
\cite{Wintz} \\
BR(\tau \ra \mu \gamma) \, < 1.1 \times \, 10^{-6} \, && \, \,
\cite{CLEO}
\end{eqnarray}
however, 
projects are currently under way to improve these upper limits
significantly, especially in 
intense $\mu$ sources that might
improve especially 
the upper limits on $\mu \ra e$ transitions 
by several orders of magnitude~\cite{mufact}.
This indicates that it is of fundamental importance
to understand the magnitude of the effects that 
one might expect, in association with neutrino oscillations.

\section{Neutrino threshold effects}

As we have already mentioned in the introduction,
the running of couplings from the unification scale,
$M_{GUT}$, to low
energies, is modified by neutrino thresholds.
The Dirac neutrino Yukawa coupling, 
$\lambda_N$, runs until the scale $M_N$. Subsequently it
decouples and the quantity that runs is the effective neutrino
operator $m_{eff}$.

In order to understand the renormalization effects
due to a non-zero $\lambda_N$
between $M_{GUT}$ and $M_N$, it is easier to start with
the small-$\tan\beta$ regime of a supersymmetric
theory, where only the top and the Dirac 
neutrino Yukawa couplings contribute
in a relevant way. 
The effect of the neutrino coupling to the gauge interactions
is smaller than its effect 
to the Yukawa couplings, since it is only at two loop
that $\lambda_N$ enters in the running of $\alpha_i$.
In a diagonal basis \cite{VS}, the renormalisation
group equations of the Yukawa couplings take the following form: 
 \bea
 16\pi^2 \frac{d}{dt} \lambda_t&=
 & \left(
 6 \lambda_t^2  + \lambda_N^2
   - G_U\right)  \lambda_t \nonumber \\
 16\pi^2 \frac{d}{dt} \lambda_N&=& \left(
  4\lambda_N^2  + 3 \lambda_t^2
   - G_N \right) \lambda_N \nonumber   \\
 16\pi^2 \frac{d}{dt} \lambda_b &=
 & \left(\lambda_t^2 - G_D \right) \lambda_b \nonumber \\
 16\pi^2 \frac{d}{dt} \lambda_{\tau}&=&\left( \lambda_N^2
  - G_E \right) \lambda_{\tau}
 \label{eq:rg4}
 \eea
where $\lambda_\alpha: \alpha=t,b, \tau ,N$, represent the
third-generation Dirac Yukawa couplings for the up and down quarks,
charged
lepton and neutrinos, respectively, 
and the $G_{\alpha} \equiv \sum_{i=1}^3c_{\alpha}^ig_i(t)^2$ are
functions that depend on the  gauge couplings, with the
coefficients $c_{\alpha}^i$ given in~\cite{VS}. 
In terms of
the various Yukawa couplings 
$\lambda_{t_0}$, $\lambda_{N_0}$,$\lambda_{b_0},{\lambda_{\tau_0}}$,
at the  unification scale, we can derive simple
expressions which indicate how neutrinos
affect the rest of the Yukawa couplings of the theory.
Indeed~\cite{LLR}:
\bea
 \lambda_t(t)&=&\gamma_U(t)\lambda_{t_0}\xi_t^6\xi_N ~~~~~ 
\lambda_N(t)=\gamma_N(t)\lambda_{t_0}\xi_t^3\xi_N^4\\
 \lambda_b(t)&=&\gamma_D(t)\lambda_{b_0}\xi_t ~~~~~~~~~ 
\lambda_{\tau}(t)=\gamma_E(t)\lambda_{\tau_0}\xi_N
\\
 \gamma_\alpha(t)&=&  \exp\left ({\frac{1}{16\pi^2}\int_{t_0}^t
  G_\alpha(t) \,dt} \right ) = \prod_{j=1}^3 \left( 
\frac{\alpha_{j,0}}{\alpha_j}
 \right)^{c_\alpha^j/2b_j} \\
 \xi_i&=& \exp
\left ({\frac{1}{16\pi^2}\int_{t_0}^t \lambda^2_{i}dt} \right )
 \eea
As noted, these results are valid for small $\tan\beta$.
For large $\tan\beta$, the bottom and tau Yukawa
couplings start playing an important role
and the complete form of the renormalisation
group equations is given in \cite{rge}.

Below the right-handed Majorana mass scale,
where $m_{eff}$ is formed,
${\lambda}_N$ decouples 
from the renormalisation group equations.
However, the 
effective neutrino mass operator will  be a
running quantity. 
For a generic $\tan\beta$ 
\bea
\frac{1}{m_{eff}^{ij}}\frac{d}{d t}m_{eff}^{ij}&=& 
\frac{1}{8\pi^2}\left( -c_i g_i^2 + 3 \lambda_t^2 
+ \frac{1}{2} ( \lambda_{i}^2 + \lambda_j^2)   \right) 
\label{RGmeff}
\eea
where $i,j$ are lepton flavour indices,
already indicating that large Yukawa terms,
which lower the effective couplings,
have a larger effect on $m_{eff}^{33}$ than
on the other entries. Finally,
the neutrino mixing angle relevant to the
atmospheric neutrino deficit, $\theta_{23}$,
is also a running quantity,
given by \cite{Bab,ELLN2}:
\bea
    16\p^2 {d\over dt}{\sin^2 2\th_{23}} & =& 2\sin^2 2\th_{23}
        (1-2 \sin^2 \th_{23}) \nonumber \\
& &  (\lambda_{\tau}^2-\lambda_{\mu}^2) ~{\k^{33}+\k^{22}\over
\k^{33}-\k^{22}}   
\label{rune}
\eea
where the initial conditions for the running
from $M_N$ down to low energies are determined
by the running of couplings between $M_{GUT}$ and
$M_N$.

\section{Renormalisation of the neutrino mass operator
and stability properties of neutrino textures}

From eq.(\ref{RGmeff}), we already see that 
the neutrino masses  will in fact 
vary non-trivially with the energy.
Given the very small mass differences
that are required for solutions to the solar and
the atmospheric neutrino deficits, it is natural to
ask whether a Super-Kamiokande-friendly
texture at the GUT scale is still a solution at
low energies. 

It is convenient for the subsequent discussion to
define the integrals
\bea
 I_g& =& \exp[\frac{1}{8\pi^2}\int_{t_0}^t(-c_i g_i^2 dt)]\\
I_t &=& \exp[\frac{1}{8\pi^2}\int_{t_0}^t  \lambda_t^2 dt]\\
    I_{i} &=& \exp[\frac 1{8\pi^2}\int_{t_0}^t \lambda_{i}^2 dt],
~~i = e,\mu,\tau
\eea
Simple integration of (\ref{RGmeff}) yields
\bea
\frac{m_{eff}^{ij}}{m_{eff,0}^{ij}}&=& 
\exp\left\{ \frac{1}{8\pi^2}\int_{t_0}^t
             \left(-c_i g_i^2 + 3 \lambda_t^2 
+ \frac{1}{2} ( \lambda_{i}^2 + \lambda_j^2)
\right)\right\}
\nonumber\\
    &=& I_g\cdot I_t \cdot 
 \sqrt{I_{i}} \cdot  \sqrt{I_{j}}
\label{RGis}
\eea
where the initial conditions are denoted by $ m_{eff,0}^{ij}$.
As we have already mentioned, these conditions are defined at
$M_N$, the scale where the neutrino Dirac coupling $\lambda_N$ decouples from
the renormalisation-group equations. 

Using (\ref{RGis}), we see that an initial texture $m_{eff,0}^{ij}$
at $M_N$ is modified to become~\cite{EL}
\bea
m_{eff}  \propto
\left (
\begin{array}{ccc}
m_{eff,0}^{11} ~I_e & m_{eff,0}^{12} 
~\sqrt{I_\mu}  ~\sqrt{I_e} 
& m_{eff,0}^{13} ~\sqrt{I_e}  ~\sqrt{I_\tau} \\
 & & \\
m_{eff,0}^{21} ~\sqrt{I_\mu}  ~\sqrt{I_e} 
& m_{eff,0}^{22} ~I_\mu & m_{eff,0}^{23} 
~\sqrt{I_\mu}  ~\sqrt{I_\tau} \\
 & & \\
m_{eff,0}^{31} ~\sqrt{I_e}  ~\sqrt{I_\tau}
 & m_{eff,0}^{32} 
~\sqrt{I_\mu}  ~\sqrt{I_\tau} & 
m_{eff,0}^{33}  ~I_\tau
\end{array}
\right) 
\label{factor}
\eea 
at $m_{SUSY}$
\footnote{For small $\tan\beta$, ignoring as a first
approximation all the Yukawa couplings except the top one,
an initial texture 
$m_{eff}(M_N)^{ij}$
at $M_N$ becomes at  a lower scale
$ m_{eff}  \propto
I_g \cdot I_t \cdot  m_{eff}(M_N) $. 
}.

Even before performing a complete numerical analysis,
we can make several observations \cite{EL}:

$\bullet$
First, note that the relative structure of
$m_{eff}$ is only modified by the charged-lepton
Yukawa couplings.
On the contrary, the top and gauge couplings give
only an overall scaling factor.

$\bullet$
Because of the factorization in
(\ref{factor}), although the individual masses and 
mixings get modified, any mass matrix that is
singular with a vanishing determinant -
leading to a zero mass eigenvalue -
remains so at the one-loop level.

$\bullet$
The Yukawa renormalization factors $I_i$ are less
than unity, and lead to the mass ordering
$m_{\nu_e} > m_{\nu_\mu} > m_{\nu_\tau}$, if we start with
exactly degenerate neutrinos at $M_{GUT}$.

$\bullet$
For values of $I_{\tau,\mu}$ 
different from unity,
the renormalization effects can be significant
even for the 
light-generation sector, since,
very small mass differences are
required for addressing the solar neutrino problem.

In order to quantify the renormalization effects on the 
physical neutrino masses,
we can start with a texture which, at the GUT scale, leads to
three {\em exactly degenerate} neutrinos, such as \cite{GGL}:
\bea
{m_{eff}} \propto \,
\left (
\begin{array}{ccc}
0 &  {1\over\sqrt2} &  {1\over\sqrt2}  \\
& & \\
{1\over\sqrt2} &  {1\over2} &  -{1\over2}   \\
& & \\
{1\over\sqrt2}  &  -{1\over2}  
 &  {1\over2}  
\end{array}
\right)
\label{GGL}
\eea
 with 
scaled eigenvalues 1,-1,1 and calculate the respective values
at low energies \cite{EL}.
We take as
illustrative initial conditions $\alpha_{GUT}^{-1} = 25.64$, 
$M_{GUT} =  1.1 \cdot 10^{16}$~GeV and
$m_{SUSY} = 1$ TeV.
We also choose $\lambda_b / \lambda_\tau$ 
such that an intermediate scale $M_N = 10^{13}$~GeV is
consistent with the observed pattern of fermion masses.
The values of $I_\tau$ and $I_\mu$ that we find \cite{EL} 
are presented in Table~1 and can be used to
estimate the effects on the 
neutrino eigenvalues, mixings
and mass differences,
as shown in the last three columns of Table~1, and in Fig.~1.

\begin{table}[tbp]
\begin{center}
\begin{small}
\begin{tabular}{|c|c|c|c|c|c|} \hline \hline
$\lambda_\tau$  & 
$I_\tau$ & 
$I_\mu$  &
$m_{3}$  &
$m_{2}$ &
$m_{1}$  \\
\hline
         3.0 &   0.826  & 0.9955  & 0.866 & -0.952 &  0.997
\\ \hline
         1.2 &  0.873  & 0.9981 & 0.903 & -0.966 &  0.998 
\\ \hline
         0.48 &  0.9497   & 0.9994  & 0.962 & -0.987 & 0.9996
\\ \hline
        0.10 &  0.997   & 0.99997  & 0.9478 & -0.9993 & 0.99998 
\\ \hline
        0.013 &  0.99997   & 1.00000  & 0.99998 & -0.99999 & 1.00000
\\ \hline \hline 
\end{tabular}
\end{small}
\end{center}
\caption{
{\it  Values of $I_\tau$ and $I_\mu$, for $M_{N} = 10^{13}$~GeV
and different choices of $\lambda_\tau$. Also tabulated are the three
renormalized mass eigenvalues calculated for three exactly degenerate
neutrinos for the unrenormalised texture.
}}
\end{table}

\begin{figure}[tbp]
\vspace*{-3.4 cm}
\centerline{\epsfig{figure=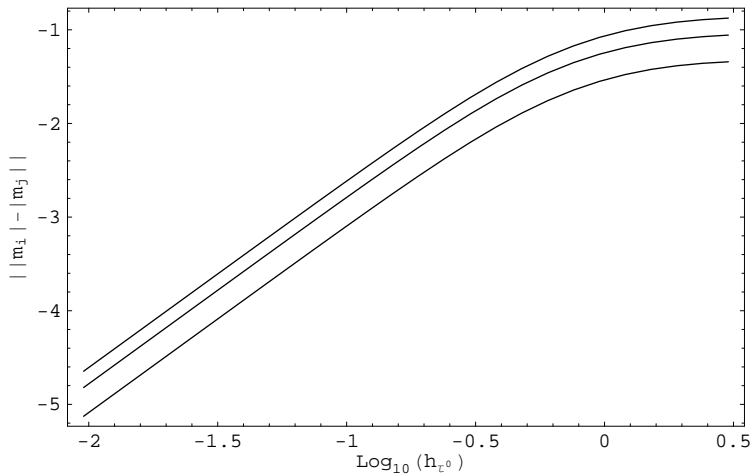,width=1.2\textwidth,clip=}}
\vspace*{-8.0 cm}
\caption{
{\it 
Renormalization of $m_{eff}$ eigenvalues for different 
initial values of $\lambda_\tau$
corresponding to values of $\tan\beta$ in the range 1 to 58,
assuming three exactly degenerate neutrinos
and $M_N = 10^{13}$~GeV.
We see that the vacuum-oscillation scenario is never accommodated.}}
\end{figure}

We see that the 
renormalization effects on the neutrino-mass
eigenvalues  are significant and spoil the neutrino
degeneracy. It is apparent from Table~1 and Fig.~1 
that the breaking of
the neutrino-mass degeneracy in this model is unacceptable
even for small $\tan\beta$ \cite{EL}.
However, we should add that
ways to stabilise the neutrino textures have been proposed:
For instance, one can 
have textures that owing to a symmetry 
are already non-degenerate at the high scales of the theory
\cite{BRSt}. Moreover, it may be that the
structure of the Dirac neutrino mass matrices 
stabilises the textures \cite{EC}, and examples
where this can happen have been proposed.

In addition, as indicated by eq.(\ref{rune}),
even the mixing angle
may significantly change from the
GUT scale to low energies
(i) if $\lambda_{\tau}$ is large, and (ii)
if the diagonal entries of $m_{eff}$ are 
close in magnitude.
To quantify this statement analytically,
we may integrate the differential
equations for the diagonal elements
of  the effective neutrino mass matrix \cite{ELLN2},
yielding the result  
\bea
   \frac{m_{eff}^{33}+m_{eff}^{22}}{m_{eff}^{33}-m_{eff}^{22}}& =&
    \frac{m_{eff,0}^{33} I_{\tau} +m_{eff,0}^{22}}
{m_{eff,0}^{33} I_{\tau}-m_{eff,0}^{22}} \equiv  f(I_{\tau})
\nonumber 
\eea
Because of the running of 
the $\tau$ Yukawa coupling being larger than those
for the other flavours of charged leptons,
$m_{eff}^{33}$ decreases
more rapidly than
$m_{eff}^{22}$. Then, if one starts with
 $m_{eff}^{22} < m_{eff}^{33}$ but both still
relatively close in magnitude, for a sufficiently large $I_\tau$,
at a given scale we obtain 
$m_{eff}^{22} = m_{eff}^{33}$ 
and the mixing angle becomes {\em maximal}.
The larger $\lambda_{\tau 0}$,  the earlier the entries
may become equal.
The exact scale where the mixing angle is maximal
is given by the relation
\bea
I_{\tau} = \frac{m_{eff,0}^{22}}{m_{eff,0}^{33}}
\eea
After reaching the maximal angle at some
intermediate scale, the running of
$\lambda_{\tau}$ results in
$ {m_{eff,0}^{33}}  < {m_{eff,0}^{22}}  $,
changing the sign of
$f(I_{\tau})$ and thus resulting in a decrease of
the mixing. 
In order, therefore, for a texture of this type to be
viable, there needs to be a balance between
the magnitudes of $\lambda_{\tau}$ and
$m_{eff}^{33}-m_{eff}^{22}$ at the GUT
scale \footnote{For an alternative
approach to the problem and a detailed discussion
of fixed points for neutrino mixing angles,
see also \cite{Poko}.}.

Moreover, there might be additional intrinsic
instabilities on the
neutrino mixing. To see this, let us 
discuss the renormalization of the neutrino 
mixing angles, considering
a perturbation $\epsilon$ from the texture in
eq.(\ref{GGL}):
\bea
{m_{eff}'} \propto \,
\left (
\begin{array}{ccc}
0 &  {1\over\sqrt2} &  {1\over\sqrt2} 
(1+{\epsilon \over 2}) \\
& & \\
{1\over\sqrt2} &  {1\over2} &  -{1\over2} (1+{\epsilon \over 2})  \\
& & \\
{1\over\sqrt2} (1+{\epsilon \over 2}) &  -{1\over2}  
(1+{\epsilon \over 2}) &  {1\over2} (1+{\epsilon }) 
\end{array}
\right)
\label{GGtextureP}
\eea
Here, $\epsilon$ is a small quantity,
which might arise from renormalisation group running
or from some other higher-order effects
such as higher-dimensional non-renormalizable operators.
This perturbation lifts the degeneracy of the eigenvalues,
which are now given by
\bea
1, ~~ -1 - {\epsilon \over 4}, ~~1 + {3 \epsilon \over 4}
\nonumber 
\eea
To this order, the eigenvectors
are independent of $\epsilon$ and given by
\bea
V_1 = \left (
\begin{array}{c}
{1 \over \sqrt{3}} \\ \sqrt{ \frac{2}{3} } \\ 0 
\end{array}
\right ),~~
V_2 = \left (
\begin{array}{c}
{1 \over \sqrt{2}} \\ -{1 \over 2}  \\ -{1 \over 2} 
\end{array}
\right ), ~~ V_3 = 
\left (
\begin{array}{c}
{1 \over \sqrt{6}} \\ -{1 \over {2 \sqrt{3}}}  \\ {\sqrt{3}\over {2}} \\
\end{array}
\right )
\label{vec1}
\eea
so that the mixing expected in this
type of texture does not depend on $\epsilon$,
as long as it is non-zero.
The vectors (\ref{vec1}) are also eigenvectors  of the
unrenormalised texture (with $\epsilon = 0$).

Let us now go back to the unperturbed texture.
Since the latter has two exactly degenerate
eigenvalues, there is
arbitrariness in the choice of eigenvectors:
the vectors corresponding to the
two degenerate eigenvalues are not linearly independent,
and can be rotated to different
linear combinations, which 
still obey the orthogonality conditions.
One example is the choice
\bea
V_1 & = & \frac{1}{\sqrt{3}} V_1'  
+ \sqrt{\frac{2}{3}} V_3'  \nonumber \\
V_3 & = & \frac{1}{\sqrt{3}} V_3' 
-\sqrt{\frac{2}{3}} V_1'  \nonumber 
\eea
which gives
\bea
V_1' = \left (
\begin{array}{c}
0 \\ \frac{1}{\sqrt{2}} \\ -\frac{1}{\sqrt{2}} 
\end{array}
\right ),~~
V_2' = \left (
\begin{array}{c}
{1 \over \sqrt{2}} \\ -{1 \over 2}  \\ -{1 \over 2} 
\end{array}
\right ), ~~ 
V_3' = \left (
\begin{array}{c}
\frac{1}{\sqrt{2}} \\ \frac{1}{2} \\
\frac{1}{2} 
\end{array}
\right )
\label{vec2}
\eea
corresponding to bimaximal mixing: $\phi_1 = {\pi \over 4}$,
$\phi_2 = 0$ and $\phi_3 = {\pi \over 4}$. However, one cannot
in general expect this latter combination of eigenvectors to
be stable when the degenerate texture is perturbed,
and the above analysis shows that, indeed, it is not.
On the contrary,
it is the direction given by (\ref{vec1}) that is stable.
Moreover, the absence of the parameter $\epsilon$ in the eigenvectors
indicates that for this texture
we may expect only minor modifications
in the mixing, for $\tan\beta$ between 1 and 60.

\section{
Neutrino threshold effects and Yukawa
unification
}

From eqs.(\ref{eq:rg4}), we see that the 
Dirac neutrino Yukawa coupling,$\lambda_N$, 
will modify  the
ratio of $\lambda_b/\lambda_\tau$ (since the top
Yukawa is close to a fixed point, the effects in it
are relatively small).
Using (\ref{eq:rg4}) we find that:
 \begin{equation}
 \lambda_{b}(t_N)=\rho
 \xi_t\frac{\gamma_D}{\gamma_E}\lambda_{\tau}(t_N), ~~~
\rho=\frac{\lambda_{b_0}}{\lambda_{\tau_0}\xi_N}
 \end{equation}
For $b$--$\tau$ unification at $M_{GUT}$, 
  $\lambda_{\tau_0} =\lambda_{b_0}$. In the absence
of a right-handed neutrino, $\xi_N \equiv 1$, 
 $\rho =1 $ and $m_b$  at low
 energies is correctly predicted.
 In the presence of $\nu_R$, however,  $\lambda_{\tau_0}
 =\lambda_{b_0}$ at the GUT scale implies that 
$\rho \neq 1$  (since $\xi_N<1$). 
To restore $\rho$ to unity, a deviation from bottom--tau unification
is required. For example,
for $M_N \approx 10^{13}~{\rm GeV}$ and 
$\lambda_{t_0} \ge 1$, it turns out that
 $\xi(t_N)\approx 0.89$. This corresponds to an
approximate  $10\%$ deviation from the $\tau$--$b$ equality at the
GUT scale, in agreement with the numerical results.

For large $\tan\beta$, even ignoring
large corrections to $m_b$ from superparticle loops \cite{Hal,CW},
the effect
of the heavy neutrino scale is much smaller, since now
the bottom Yukawa coupling also runs to a fixed point
\cite{CW} \footnote{
For large $\tan\beta$ and $\lambda_b\approx \lambda_{t}$,
the product and ratio of the top and bottom
couplings can be simply expressed as
$ \lambda_t \lambda_b \approx
\frac{{8\pi^2}\gamma_Q \gamma_D}{{7}\int \gamma_Q^2 d\,t},
~\frac{\lambda_t^2}{\lambda_b^2}\approx \frac{\gamma_Q^2}{\gamma_D^2}$ 
 \cite{FLL},
indicating that there is an approximate,
model-independent  prediction for both couplings
at the low-energy scale.}.

We can confirm these results by a numerical analysis,
for neutrino parameters that are compatible with Super-Kamiokande
\cite{CELW}. To do so, in \cite{CELW} we choose 
a scale $M_{GUT} \simeq 1.5 \times 10^{16}$ GeV,
for which approximate unification of the three gauge couplings
holds. We also choose a soft supersymmetry
breaking scale of the order of 1 TeV,  $\alpha_3(M_Z) \simeq 0.118$,
$m_{top}=175$ GeV and
$m_{bottom}^{pole}= 4.8$ GeV.
We then plot the ratio $m_{\tau}/m_b(M_{GUT})$, as a
function of $M_N$, for fixed values of $\tan\beta$ \cite{CELW}. 
This is shown in Fig.~2, for a neutrino
mass value  $m_{\nu} = 0.03$ eV.
In the figure, the lines are truncated when the value of $M_N$ is
such, that the neutrino Yukawa coupling enters the non-perturbative
regime.

\begin{figure}[t]  
\centerline{
\psfig{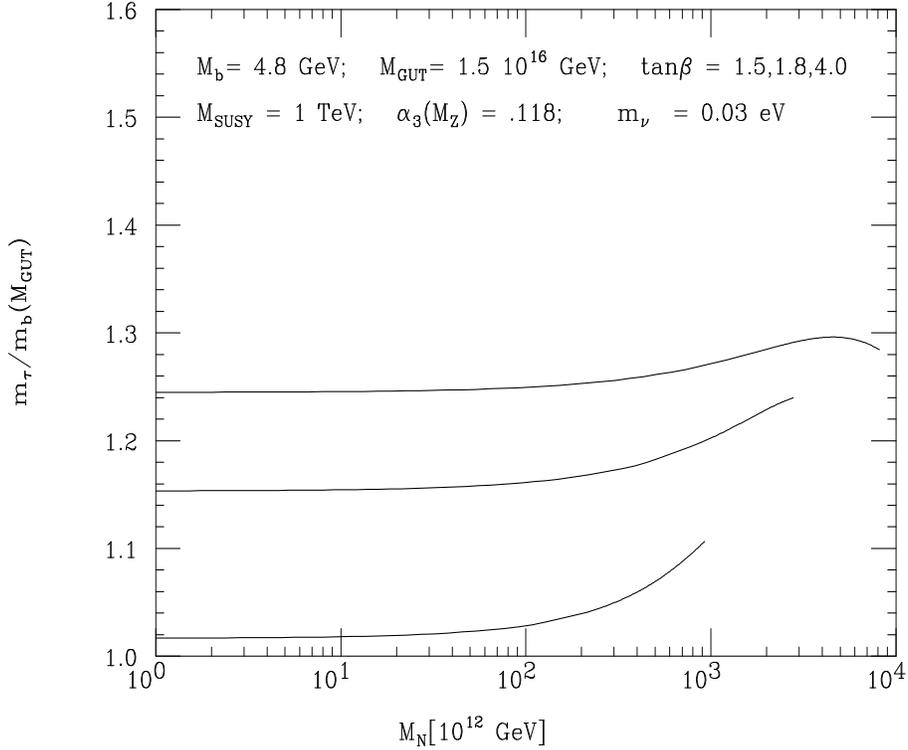}}
%bbllx=7.5cm,%
%bblly=5.cm,bburx=16.5cm,bbury=14.cm}}
\caption
{\it The ratio
$m_\tau/m_b(M_{GUT})$ as a function of
$M_N$, with the choice  
$m_{\nu}$ = 0.03 eV and
for different values of 
$\tan\beta$: from bottom to top, $\tan\beta$ = 1.5, 1.8 and 4, respectively. }
\end{figure}

We can then make the following observations:

(i) For small $\lambda_N$ 
(small $M_N$ in the see-saw model)
the appearance of
the neutrino masses does not play a major role.
For small $\tan\beta$,
in the region of the top infrared fixed-point, we 
obtain $b-\tau$ unification;
when $\tan\beta$ increases, the expected deviation 
from $b-\tau$ unification is seen.

(ii) As $\lambda_N$ becomes larger for fixed $\tan\beta$
(large $M_N$),
the neutrino coupling lowers $\lambda_{\tau}$ with
respect to $\lambda_b$; thus,
to obtain the correct value of $m_b/m_\tau$
at low energies, we need to
start with lower $\lambda_b/\lambda_\tau(M_{GUT})$.

(iii) As $\lambda_N$ increases, 
$M_N$ gets close to the GUT scale and $\ln(M_N/M_{GUT})$
decreases the magnitude of the effects. 
This explains the presence of a peak for
$\tan\beta = 4$.
For the other values of $\tan\beta$ 
the Dirac neutrino 
coupling is so large
that $\lambda_N$ enters the non-perturbative regime before
this peak is reached.

Given these results, it is natural to ask if
models with $b$--$\tau$
equality and large Dirac neutrino coupling
at $M_{GUT}$ may be consistent with the 
required neutrino masses in the small $\tan\beta$ regime.
To answer this, we need to remember that
the $b$--$\tau$ equality at the GUT scale refers to the
$(3,3)$ entries of the charged-lepton and
down-quark mass matrices
(denoted by $(m^{diag}_\ell)_{33}$ and $(m^{diag}_{down})_{33}$ 
respectively), while the detailed structure of the
mass matrices is not predicted by the grand unified
group itself.
It is then possible to assume mass textures,
such that, after the diagonalisation at   the
GUT scale, the $(m^{diag}_\ell)_{33}$ and $(m^{diag}_{down})_{33}$
entries are no longer equal \cite{LLR}.

To understand the effect, we consider
a $2 \times 2$ example, and  assume that the off-diagonal terms
in the down-quark mass matrix are small compared to
the (33) element, whereas this is not the case for the
charged-lepton mass matrix. In this case, one can approximate the 
down-quark and charged-lepton mass matrices at the GUT scale by
\bea
\label{d0e0}
m_{down}^0 = A 
 \left (
 \begin{array}{cc}
c & 0 \\
0 & 1 
 \end{array}
 \right), ~~~ m_{\ell}^0 = A 
 \left (
 \begin{array}{cc}
x^2 & x \\
x & 1 
 \end{array}
 \right),
\eea
where $A$ may be identified with $m_b(M_{GUT})$, the 
bottom quark mass at the scale $M_{GUT}$.
At low energies, the eigenmasses are obtained by
diagonalising the renormalized Yukawa matrices;
this is equivalent to
diagonalising the quark and charged-lepton Yukawa matrices
at the GUT scale, and then evolving the eigenstates and
the mixing angles separately. In this way,
we see that the trace of the charged-lepton mass matrix,
which gives the higher eigenvalue,
is not 1, but $1+x^2$, and therefore 
the effective $\lambda_b$ and $\lambda_\tau$ 
are not equal after diagonalization.

In cases where large deviations from $b-\tau$ unification
is found, this unification can be restored by introducing
large mixing in the charged-lepton sector.
Then the mixing in the neutrino sector, 
is also calculable. Both mixings appear
on the tables 2 and 3 respectively, indicating that
Yukawa unification can be a useful independent probe of
neutrino and charged-lepton textures.

\begin{table}[tbp]
\begin{center}
\begin{tabular}{|c|c|c|c|c|c|c|c|c|}
\hline
$M_N[10^{13}$ GeV] & 1 & 10 & 20 & 50 & 70 & 150 & 250 & 400 \cr
\hline
$\tan\beta = 1.5$ &0.13 &0.15 & 0.17 & 0.21 &0.23 & & &\cr
\hline
$\tan\beta = 1.8$  &0.39 &0.40 &0.40&0.41&0.42 &0.43 & 0.44&\cr
\hline
$\tan\beta = 4.0$&0.50 &0.50& 0.50&0.50&0.50 &0.51 &0.52 & 0.52\cr
\hline
\end{tabular}
\end{center}
\caption{\it Values of charged lepton $\mu-\tau$ mixing
leading to $b-\tau$
Yukawa coupling unification for $m_{\nu}= 0.03$ eV,
for different choices of $\tan\beta$ and $M_N$.}
\label{tab:x003}
\end{table}

\begin{table}[tbp]
\begin{center}
\begin{tabular}{|c|c|c|c|c|c|c|c|c|}
\hline
$M_N[10^{13}$ GeV] & 1 & 10 & 20 & 50 & 70 & 150 & 250 & 400 \cr
\hline
$\tan\beta = 1.5$ &-0.77 &-0.73 & -0.69 &-0.62& -0.58 & & &\cr
\hline
$\tan\beta = 1.8$  &-0.44 &-0.43&-0.42&-0.40&-0.39 &-0.37 &-0.35 &\cr
\hline
$\tan\beta = 4.0$&-0.34 &-0.33&-0.33&-0.32&-0.32 &-0.31 &-0.30 &-0.29\cr
\hline
\end{tabular}
\end{center}
\caption{\it Values of neutrino $\mu-\tau$ mixing
leading to $b-\tau$ Yukawa coupling
unification for $m_{\nu}= 0.03$ eV,
for different choices of $\tan\beta$ and $M_N$.}
\label{tab:y003}
\end{table}

\section{Lepton-flavour-violating rare processes}

In the Standard Model (SM) with massive
neutrinos, $\mu \rightarrow e \gamma$ is mediated by 
diagrams of the type~\cite{neutrinoLFVns}:
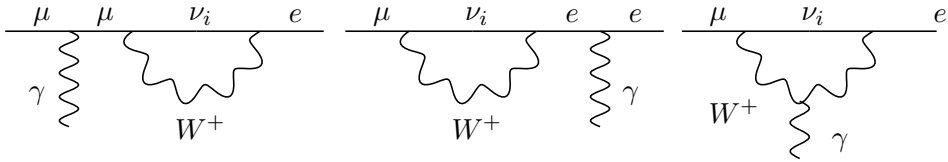
\begin{figure}[h]
{\unitlength=1.2 pt
\SetScale{1.2}
\SetWidth{0.5}      
\begin{picture}(100,100)(0,0)
\Line(0,60)(20,60)
\Line(20,60)(40,60)
\Line(40,60)(60,60)
\Line(60,60)(80,60)
\Line(80,60)(100,60)
\PhotonArc(60,60)(20,180,360){3}{5}
\Photon(20,60)(20,30){3}{4}
\Text(10,65)[]{ $\mu$}
\Text(30,65)[]{ $\mu$}
\Text(90,65)[]{ $e$}
\Text(10,40)[]{$ \gamma$}
\Text(60,65)[]{ $\nu_i$}
\Text(60,30)[]{ $W^+$}
\end{picture}
\hspace*{-0.8 cm}
\begin{picture}(100,100)(0,0)
\Line(20,60)(40,60)
\Line(40,60)(60,60)
\Line(60,60)(80,60)
\Line(80,60)(100,60)
\Line(100,60)(120,60)
\PhotonArc(60,60)(20,180,360){3}{5}
\Photon(100,60)(100,30){3}{4}
\Text(30,65)[]{ $\mu$}
\Text(110,65)[]{ $e$}
\Text(90,65)[]{ $e$}
\Text(110,40)[]{$ \gamma$}
\Text(60,65)[]{ $\nu_i$}
\Text(60,30)[]{ $W^+$}
\end{picture}
\hspace*{-0.0 cm}
\begin{picture}(100,100)(0,0)
\Line(20,60)(40,60)
\Line(40,60)(60,60)
\Line(60,60)(80,60)
\Line(80,60)(100,60)
\PhotonArc(60,60)(20,180,360){3}{5}
\Photon(57,38)(57,20){3}{2}
\Text(30,65)[]{ $\mu$}
\Text(100,65)[]{ $e$}
\Text(70,25)[]{$ \gamma$}
\Text(60,65)[]{ $\nu_i$}
\Text(35,35)[]{ $W^+$}
\end{picture}
}
\caption{Minimal Standard Model plus massive neutrino
contributions to $\mu \rightarrow e \gamma$
}
\end{figure}

The decay rate for these processes is proportional
to the neutrino mass square difference, scaling as
$\Gamma \propto  
\frac{(m_2^2-m_1^2)}{m_W^2}~\sin^2\theta \cos^2\theta$.
For $\delta m_{12}^2$ in the range indicated by the
neutrino data,
the branching ratio for this decay is
$\leq 10^{-50}$, and thus too small to observe.
The same is true for 
the rest of the flavour-violating processes, such as
$\tau \rightarrow \mu \gamma$, $\mu \rightarrow 3 e$ and
$\mu-e$ conversion in nuclei.

However the situation is vastly different in supersymmetry
~\cite{neutrinoLFVs,a}, where in the presence of 
$\tilde{\mu}$-$\tilde{e}$ 
($\tilde{\nu}_\mu$-$\tilde{\nu}_{e}$)
mixing, one can generate the diagrams of Fig. 4:

\begin{figure}[h]
\hspace*{-1.0 cm}
{\unitlength=0.6 pt
\SetScale{0.6}
\SetWidth{1.0}
\begin{picture}(350,350)(0,0)
\Line(50,230)(130,230)
\DashLine(130,230)(230,230){3}
\Line(290,230)(230,230)
\CArc(180,230)(50,0,180)
\Photon(180,230)(180,170){3}{4}
\Text(100,215)[]{ $\mu$}
\Text(260,215)[]{ $e$}
\Text(195,185)[]{ $\gamma$}
\Text(150,213)[]{ $\tilde{\ell}_i$}
\Text(162,300)[]{ ${\tilde \chi}^0$}
\end{picture}    
\hspace*{-2.0 cm}  
\begin{picture}(350,350)(0,0)
\Line(50,230)(130,230)
\DashLine(130,230)(230,230){3}
\Line(290,230)(230,230)
\CArc(180,230)(50,0,180)
\Photon(220,259.95)(260,309.55){4}{4}
\Text(100,215)[]{ $\mu$}
\Text(260,215)[]{ $e$}
\Text(260,285)[]{ $\gamma$}
\Text(180,213)[]{ $\tilde{\nu}_i$}
\Text(162,300)[]{ ${\tilde \chi}^-$}
\end{picture}
}
\vspace*{-3 cm}
\caption{Supersymmetric contributions to
$\mu \rightarrow e \gamma$}
\end{figure}
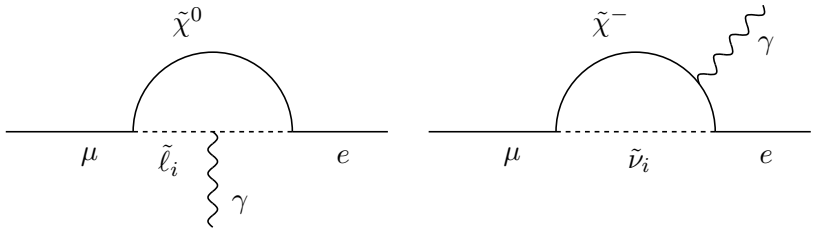

Since the   fermion in the loop is now a
neutralino/chargino instead of a neutrino as in the
previous case
( with $m_{\tilde{\chi}^0},m_{\tilde{\chi}^\pm} \gg m_{\nu}$),
much larger rates are expected.
The magnitude of the rates depends on
the masses and mixings of superparticles.
For non-universality at $M_{GUT}$,
large rates are in general predicted.
However,  even if
at ${M_{GUT}}$
\[
m_{\tilde{\ell},\tilde{\nu}} \propto \left (
\begin{array}{ccc}
1 & 0 & 0 \\
0 & 1 & 0 \\
0 & 0 & 1
\end{array} 
\right )
\]
renormalisation effects
of the Minimal Supersymmetric Standard Model
(MSSM) with right-handed neutrinos
will spoil this diagonal form ~\cite{neutrinoLFVs,a}
to give 
\[
m_{\tilde{\ell},\tilde{\nu}} \propto \left (
\begin{array}{ccc}
1 & \star & \star \\
\star & 1 & \star \\
\star & \star & 1
\end{array}
\right )
\]
Indeed,  the Dirac neutrino and charged-lepton Yukawa
couplings cannot, in general, be diagonalized simultaneously;
since both these sets of lepton Yukawa
couplings appear 
in the renormalisation-group equations, neither
the lepton Yukawa
matrices nor the slepton mass matrices can be simultaneously
diagonalized at low energies either.
In the basis where $m_{\ell}$
is diagonal, the slepton-mass matrix acquires 
non-diagonal contributions from renormalization at scales below
$M_{GUT}$, of the form \cite{neutrinoLFVs}:
\bea
\delta{m}_{\tilde{\ell}}^2\propto \frac 1{16\pi^2} (3 + a^2)
\ln\frac{M_{GUT}}{M_N}\lambda_N^{\dagger} \lambda_N m_{3/2}^2,
\label{offdiagonal}
\eea
where $a$ is  related to the trilinear mass parameter, 
$A_\ell= a m_{3/2}$ and
$m_{3/2}^2$ is the common value of the scalar masses at the GUT
scale. 

We stress that the effects of massive neutrinos are
significant for theories with universal scalar masses at the GUT
scale, such as no-scale~\cite{Ellis:1984bm}
and gauge-mediated models~\cite{gaugmed}. In models
with non-universality at the GUT scale, excessive rates
are generically predicted.
In particular, for models with universality at $M_{GUT}$,
different predictions for the various solutions
of the solar neutrino deficit \cite{bimaximal,abel}
(with a small/large mixing angle and with eV or 
$\approx$ 0.03 eV neutrinos), predict in general 
different rates for lepton-flavour violation:
the larger the $\mu-e$ mixing
and the larger the neutrino mass scales that are required,
the larger the rates. This already indicates that
for degenerate neutrinos with bimaximal mixing,
we expect significantly larger effects than,
for instance, for hierarchical neutrinos with
a small vacuum mixing angle. Note however that,
for the just-so solutions to the solar neutrino
problem (where a $\delta m^2 \approx 10^{-10}$ eV$^2$
is required), 
the predicted rates in the case of hierarchical
neutrinos are small,
even if the (1-2) mixing is large.

In order to estimate the expected effects, 
we can calculate the
rates for rare processes \cite{GELLN}, in 
a model based on abelian flavour symmetries and
symmetric mass matrices \cite{IR}. 
For a
charged-lepton matrix with a large
(2-3) mixing in this model 
\bea
M_{\ell }  \propto  \left( 
\begin{array}{ccc}
\bar{\epsilon}^{7} & \bar{\epsilon}^{3} & \bar{\epsilon}^{7/2} \\ 
\bar{\epsilon}^{3} & \bar{\epsilon} & \bar{\epsilon}^{1/2} \\ 
\bar{\epsilon}^{7/2} & \bar{\epsilon}^{1/2} & 1
\end{array}
\right),
m^D_{\nu} \propto \left( 
\begin{array}{ccc}
{\epsilon}^{7} & {\epsilon}^{3} & {\epsilon}^{7/2} \\ 
{\epsilon}^{3} & {\epsilon} & {\epsilon}^{1/2} \\ 
{\epsilon}^{7/2} & {\epsilon}^{1/2} & 1
\end{array}
\right),
\eea
\bea
V_\ell = \left(
\begin{array}{ccc}
1 & \bar{\epsilon}^{2} & -\bar{\epsilon}^{7/2} \\
-\bar{\epsilon}^{2} & 1 & \bar{\epsilon}^{1/2} \\
\bar{\epsilon}^{7/2} & -\bar{\epsilon}^{1/2} & 1
\end{array}
\right), V_{\nu_D} = \left(
\begin{array}{ccc}
1 & \bar{\epsilon}^{4} & -\bar{\epsilon}^{7} \\
-\bar{\epsilon}^{4} & 1 & \bar{\epsilon} \\
\bar{\epsilon}^{7} & -\bar{\epsilon} & 1
\end{array}
\right) \label{Asolutions}
\eea
a small $\mu-e$ mixing is always
predicted, as a result of fixing the charged-lepton hierarchies
\cite{GGR}.

\begin{figure}[h]
\begin{minipage}[b]{8in}
\hspace*{1.5 cm}
\epsfig{file=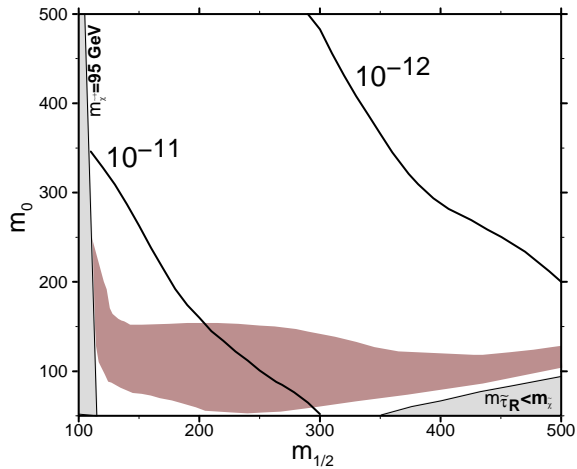,width=3in}
\hfill
\end{minipage}
\caption{\it
Contour plots in the $(m_{1/2}, m_0)$ plane for the
decay $\mu \ra e \gamma$,
assuming $\tan\beta = 3$ and $\mu < 0$.
The rates are encouraging throughout
the dark shaded region preferred by astrophysics and cosmology~\cite{ES}.}
\label{metan3}
\end{figure}

In this framework, we calculated the rates for $\mu \rightarrow e \gamma$
and $\mu-e$ conversion, which are experimentally most promising~\cite{GELLN}.
The rates depend on supersymmetric masses and mixings;
we parametrize the supersymmetric masses in terms of the universal GUT-scale
parameters $m_0$  and $m_{1/2}$, for sfermions
and gauginos respectively, and use the renormalization-group
equations of the MSSM to calculate the low-energy sparticle
masses. Other relevant free parameters of
the
MSSM are the trilinear coupling $A$,
the sign of the Higgs mixing parameter $\mu$, and
the value of $\tan\beta$.  Here we fix the value of  $A_0=- m_{1/2}$
and consider both possible signs for 
the $\mu$ parameter.
Contour plots for $\mu \ra e \gamma$ appear in Fig.~\ref{metan3}.

We observe that, as expected, the branching ratios tend to decrease
as $m_{1/2}$ and $m_0$ increase. For $\tan\beta=3$, as chosen
in the contour plot, we
predict values of $BR(\mu\ra e\gamma)$ compatible with the
experimental
bound in most of the region
where the cosmological relic density
is in the range preferred by astrophysics
(dark shaded region in plot)~\cite{GELLN}.
In contrast, if $\tan\beta \geq 10$,
acceptable $BR(\mu\ra e\gamma)$ rates are found only
for values of $m_0 \ge 400 $ GeV. 
The light shaded areas in Fig.~\ref{metan3} correspond to the 
regions of the
$(m_{1/2}, m_0)$ plane that are excluded by LEP searches for
charginos and by the requirement that the lightest
supersymmetric particle not be charged~\cite{ES}.

Let us now briefly discuss the
rare processes $\mu \ra 3e$ and $\mu \rightarrow e$ conversion on nuclei.
These decays receive contributions from
three types of Feynman diagrams.
The first are photon ``penguin'' diagrams related to
the diagrams for
$\mu \ra e \gamma$ discussed above, where now
the photon is virtual and couples to an $e^+e^-$ 
(or a quark-antiquark) pair.
A second class of diagrams is obtained
by replacing the photon line with a $Z$ boson,
and there are also box diagrams. 
In addition, all the above types of diagrams are accompanied
by their supersymmetric analogues. 
If we restrict ourselves to the 
photonic contribution, which dominates,
we have the approximate relations \cite{KO}

\begin{equation}
{\Gamma ( \mu^+ \ra e^+e^+e^-) \over \Gamma ( \mu^+ \ra e^+ \gamma)}
\approx 6 \times 10^{-3}
\label{approxratio}
\end{equation}
and
\begin{eqnarray}
 R({\mu Ti\ra e Ti})\approx  
5.6 \times 10^{-3} BR(\mu\ra e\gamma)
\label{convratio}
\end{eqnarray}
From these two processes,
 $\mu \ra e$ conversion is the most  interesting
since,  with an intense muon source, such as that
projected for a neutrino factory or a muon collider,
experiments sensitive to rates as low as $10^{-16}$
may be feasible~\cite{mufact}. 

\begin{figure}[h]
\begin{minipage}[b]{8in}
\hspace*{1.5 cm}
\epsfig{file=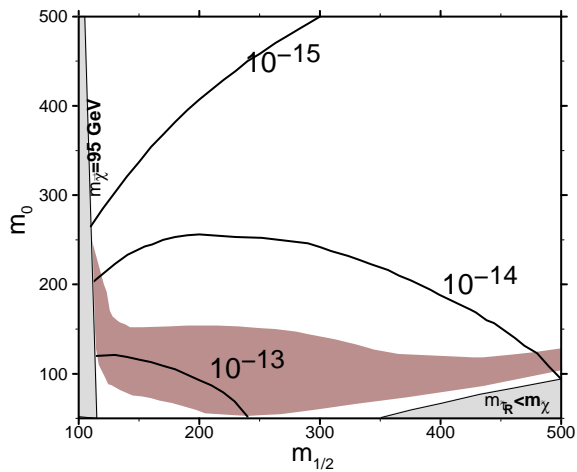,width=3 in}
\end{minipage}
\caption{\it
Contour plots in the $(m_{1/2}, m_0)$ plane for
$\mu \ra e $ conversion,
assuming $\tan\beta = 3$ 
and $\mu < 0$.
We see that the conversion rate is encouraging
throughout the dark-shaded region preferred by astrophysics and
cosmology~\cite{ES}.}
\label{come}
\end{figure}

Fig. \ref{come} displays contours of the rate for $\mu \ra e$
conversion in the $(m_0, m_{1/2})$ plane \cite{GELLN}.
We see that the former predicts a rather larger
rate, which offers good prospects for observation throughout
the region preferred by cosmology, in the next generation
of experiments, even for neutrino textures with
hierarchical neutrinos and $\mu-e$ mixing
in the small MSW region for the solar neutrino deficit.

\section{Summary}

We discussed various aspects of the renormalisation
effects of neutrino masses and interactions.
In supersymmetric extensions of the Standard Model, these
effects are important. In particular, 
for small $\tan\beta$, $b$--$\tau$ unification requires the presence of 
significant $\mu$--$\tau$ flavour mixing.  
On the other hand, for large $\tan\beta$, 
small mixing at the GUT scale may be amplified to 
maximal mixing at low energies, and vice versa. 
The eigenvalues of the neutrino mass operator are
also modified by quantum corrections;
given the very small mass differences
required to address the solar and atmospheric neutrino
deficits, several neutrino textures (especially those
with degenerate neutrinos with an eV mass scale), can be constrained
or, in certain models, even excluded.
Finally, while in the minimal scheme with the Standard Model
plus neutrino masses,
the rates for rare muon decays and $\mu-e$ conversion in nuclei
are very suppressed, this is no longer the case in supersymmetric
models. In this latter case, even for universality of soft terms
at the GUT scale,
quantum corrections due to lepton mixing induce rates that 
are very close to the current experimental bounds
and within probe in the next generation of experiments.

\vspace*{0.3 cm}
{\bf Acknowledgements:}
I would like to thank M. Carena, J. Ellis, 
M. Gomez, G.K. Leontaris, 
D.V. Nanopoulos, G.G. Ross and C.E.M. Wagner,
for very fruitful collaborations
on the study of the renormalisation effects of massive
neutrinos.


\begin{thebibliography}{99}


\bibitem{SKam}  Y. Fukuda et al., Super-Kamiokande collaboration, Phys.
Lett. B433 (1998) 9; Phys. Lett. B436 (1998) 33; Phys. Rev. Lett. 81 (1998)
1562.

\bibitem{chooz}  M. Apollonio et al., Chooz collaboration, Phys. Lett.
B420 (1998) 397.

\bibitem{MSW}  See, for example, L. Wolfenstein, Phys. Rev. D17 (1978)
20; S.P. Mikheyev and A.Yu. Smirnov, Yad. Fiz. 42 (1985) 1441 and
Sov. J. Nucl. Phys. 42 (1986) 913.

\bibitem{seesaw}  
M. Gell-Mann, P. Ramond and R. Slansky, 
{\it Proceedings of the
Stony Brook Supergravity Workshop}, New York, 1979, eds. P. Van
Nieuwenhuizen and D.~Freedman (North-Holland, Amsterdam).


\bibitem{mns}  Z. Maki, M. Nakagawa and S. Sakata, Prog. Theor. Phys. 28
(1962) 247.

\bibitem{neutrinoLFVns}
S.T. Petcov, Yad. Phys. 25 (1977) 641 and Sov. J. Nucl. Phys. 25 (1977) 340;
S.M. Bilenki, S.T. Petcov and B. Pontecorvo,
Phys. Lett. B67 (1977) 309.


\bibitem{neutrinoLFVs}
F. Borzumati and A. Masiero, Phys. Rev. Lett. 57 (1986) 961; 
J. Ellis and D.V.~Nanopoulos,  Phys. Lett. B110 (1982) 44;   
R. Barbieri and R. Gatto, Phys. Lett. B110 (1982) 211;       
L.J. Hall, V.A. Kostelecky and S. Raby,
Nucl. Phys. B267 (1986) 415;
G.K. Leontaris, K. Tamvakis and J.D. Vergados,  
Phys. Lett. B171 (1986) 412; 
J. Hisano, T. Moroi, K. Tobe and M. Yamaguchi, 
Phys. Rev. D53 (1996) 2442;
S.F. King and M. Oliveira, Phys. Rev. D60 (1999) 035003.



\bibitem{a}
R. Barbieri et al., Nucl. Phys. B445 (1995) 219;
S. Dimopoulos and D. Sutter, Nucl. Phys. B452 (1995) 496; 
M.E. G\'omez and H. Goldberg, Phys. Rev. { D53} (1996) 5244; 
A. Ilakovac and A. 
Pilaftsis, Nucl. Phys. {B 437} (1995) 491;
G.K. Leontaris and N.D. Tracas,
Phys. Lett. B431 (1998) 90;
M. G\'omez et al., Phys. Rev. D59 (1999) 116009;
J. Hisano and D. Nomura,
Phys. Rev. D59 (1999) 116005;
R. Kitano and K. Yamamoto, hep-ph/9905459;
Y. Okada and K. Okumura, hep-ph/9906446;
J.L. Feng, Y. Nir and Y. Shadmi,
hep-ph/9911370.


\bibitem{KO}
For a recent review, see Y. Kuno and Y. Okada, hep-ph/9909265.

\bibitem{Brooks} 
M.L. Brooks et al., MEGA collaboration, hep-ex/9905013.

\bibitem{Bellgardt}
U. Bellgardt et al., Nucl. Phys. B229 (1988) 1. 

\bibitem{Wintz}
P. Wintz, {\it Proceedings of the First International Symposium on Lepton
and Baryon Number Violation}, p. 534.

\bibitem{CLEO}
S.~Ahmed {et al.}, CLEO Collaboration,
hep-ex/9910060.


\bibitem{mufact}
See, for example:
W.J.~Marciano,
{\it Workshop on Physics at the first Muon Collider and at the Front End
of the Muon Collider};
{\it Prospective Study of Muon Storage Rings at CERN}, eds.
B.~Autin, A.~Blondel and J.~Ellis,
CERN Report 99-02 (1999).

\bibitem{VS}
F. Vissani and A. Yu. Smirnov, Phys. Lett. { B341} (1994) 173;
A. Brignole, H.~Murayama and R. Rattazzi, Phys. Lett.
{ B335} (1994)
345.

\bibitem{LLR}
G.K. Leontaris, S. Lola and G.G. Ross, Nucl. Phys. B454
(1995) 25;
S. Lola, hep-ph/9903203, 
{\it Proceedings of the 1998
Corfu Summer Institute on Elementary Particle Physics}, 
published in JHEP.

\bibitem{rge}
See for instance the last two papers in
\cite{neutrinoLFVs} and references therein.

\bibitem{Bab}  K. Babu, C. N. Leung and J. Pantaleone, Phys. Lett. {\ B319}
(1993) 191; P.H. Chankowski and Z. Pluciennik, 
Phys. Lett. {\ B316} (1993)
312; M. Tanimoto, Phys. Lett. B360 (1995) 41;
N. Haba, Y. Matsui, N. Okamura and M. Sigiura,
Eur. Phys. J. C10 (1999) 677 and hep-ph/9908429;
N. Haba at al., hep-ph/9911481; 
N. Haba and N. Okamura, hep-ph/9906481.

\bibitem{ELLN2}
J. Ellis et al., Eur. Phys. J. C9 (1999) 389.

\bibitem{EL}
J. Ellis and S. Lola, Phys. Lett. B458 (1999) 310.

\bibitem{GGL}
H. Georgi and S.L. Glashow, hep-ph/9808293.

\bibitem{BRSt}
R. Barbieri,
G.G. Ross and  A. Strumia, JHEP 9910 (1999) 020.

\bibitem{EC}
A. Casas, J.R. Espinosa, A. Ibarra and I. Navarro, 
Nucl. Phys. B556 (1999) 3; JHEP 9909 (1999) 015;
Nucl. Phys. B569 (2000) 82; hep-ph/991042.




\bibitem{Poko}
P.H. Chankowski, W. Krolikowski and S. Pokorski,
Phys. Lett. B473 (2000) 109.


\bibitem{Hal}
L. Hall et al., Phys. Rev. D50 (1994) 7048;

\bibitem{CW}
M. Carena et al., Nucl. Phys. B426 (1994) 269.


\bibitem{FLL}
E.G. Floratos, G.K. Leontaris and S. Lola,
Phys. Lett. B365 (1996) 149.


\bibitem{CELW}
M. Carena, J. Ellis, S. Lola and  C.E.M. Wagner,
Eur. Phys. J. C12 (2000) 507.


\bibitem{Ellis:1984bm}
J.~Ellis, C.~Kounnas and D.V.~Nanopoulos,
Nucl.\ Phys.\ {B247} (1984) 373;
J.~Ellis et al., Phys. Lett. {B134} (1984) 429.

\bibitem{gaugmed}
M. Dine and A. Nelson, Phys. Rev. D48 (1993) 1277;
M. Dine, et al.,
Phys. Rev. D53 (1996) 2658;
S. Dimopoulos, S. Thomas and J.D. Wells, 
Nucl. Phys. B488 (1997) 39;
G. Giudice and R. Rattazzi, hep-ph/9801271,
and references therein.


\bibitem{bimaximal}
V. Barger, S. Pakvasa, T.J. Weiler and K. Whisnant,
Phys. Lett. B437 (1998) 107; 
A.J. Baltz, A.S. Goldhaber 
and M. Goldhaber,
Phys. Rev. Lett. 81 (1998) 5730;
See also:
R. N. Mohapatra and S. Nussinov, 
Phys. Lett. B441 (1998) 299 and
hep-ph/9809415;
C. Jarlskog, M. Matsuda and S. Skadhauge,
hep-ph/9812282;
Y. Nomura and T. Yanagida, Phys. Rev. D59 (1999) 017303;
S.K. Kang and C.S. Kim, 
Phys. Rev. D59 (1999) 091302;
Y. L. Wu, Phys. Rev. D59 (1999) 113008;
Eur. Phys. J. C10 (1999) 491;
Int. J. Mod. Phys. A14 (1999) 4313;
C. Wetterich, 
Phys. Lett. B451 (1999) 397;
R. Barbieri, 
L.J. Hall, G.L. Kane, and G.G. Ross, hep-ph/9901228.


\bibitem{abel}
Some of the many references are:
 C. Wetterich, Nucl. Phys. { B261}
(1985) 461;
G.K.~Leontaris and D.V. Nanopoulos, Phys. Lett. B212 (1988)
327; Y.Achiman and T. Greiner, Phys. Lett. B329 (1994) 33; 
H. Dreiner et al., Nucl. Phys. B436 (1995) 461;
Y. Grossman and Y. Nir,
Nucl. Phys. B448 (1995) 30;
P. Bin\'etruy, S. Lavignac and P.
Ramond, Nucl. Phys. B477 (1996) 353; 
G.K. Leontaris, S. Lola, C. Scheich and J.D. 
Vergados, Phys.
Rev. D 53 (1996) 6381; S. Lola and J.D. Vergados, Progr. Part.
Nucl. Phys. 40 (1998) 71; 
B.C. Allanach, hep-ph/9806294;
P. Bin\'etruy et al., Nucl. Phys. B496 (1997) 3,
G.~Altarelli and F.~Feruglio,
Phys. Lett. B439 (1998) 112, 
JHEP 9811 (1998) 021,
Phys. Lett. B451 (1999) 388 and
hep-ph/9905536; 
G.~Altarelli, F.~Feruglio
and I. Masina, hep-ph/9907532;
Y. Nir and Y. Shadmi, JHEP 9905 (1999) 023;
M. Hirsch et al., hep-ph/0004115.

\bibitem{GELLN}
J. Ellis, M.E. Gomez, G.K. Leontaris, S. Lola and  D. V. Nanopoulos,
hep-ph/9911459, to appear in Eur. Phys. J. C.

\bibitem{IR}  L. Ibanez and G.G. Ross, Phys. Lett. B332 (1994) 100.

\bibitem{GGR}
S. Lola and G.G. Ross, Nucl. Phys. B553 (1999) 81.



\bibitem{ES}
J. Ellis, T. Falk and K. A. Olive, Phys. Lett. B444 (1998) 367.


\end{thebibliography}
\end{document}